\documentclass[reprint,amsmath,amssymb,floatfix,citeautoscript]{revtex4-1}
\usepackage[latin1]{inputenc}
\pdfoutput=1
\usepackage{graphicx}
\usepackage{dcolumn}
\usepackage{bm}
\usepackage[colorlinks=true,allcolors=blue]{hyperref}
\usepackage{epstopdf}
\DeclareGraphicsExtensions{.eps}
\newcommand{\Eprint}{\eprint}
\begin{document}
\title{Electron-phonon cooling in large monolayer graphene devices}
\author{Christopher B. McKitterick}
\affiliation{Departments of Physics and Applied Physics, Yale University, New Haven, Connecticut 06520, USA}
\author{Daniel E. Prober}
\affiliation{Departments of Physics and Applied Physics, Yale University, New Haven, Connecticut 06520, USA}
\author{Michael J. Rooks}
\affiliation{Yale Institute for Nanoscale and Quantum Engineering, Yale University, New Haven, Connecticut 06520, USA}
\begin{abstract}
We present thermal measurements of large area (over $1,000$~$\mu$m$^2$) monolayer graphene samples at cryogenic temperatures to study the electron-phonon thermal conductivity of graphene. By using two large samples with areas which differ by a factor of 10, we are able to clearly show the area dependence of the electron-phonon cooling. We find that, at temperatures far below the Bloch-Gruneisen temperature $T_\mathrm{BG}$, the electron-phonon cooling power is accurately described by the $T^4$ temperature dependence predicted for clean samples. Using this model, we are able to extract a value for the electron-phonon coupling constant as a function of gate voltage, and the graphene electron-lattice deformation potential. We also present results for thermal conductance at higher temperatures, above $T_\mathrm{BG}/4$, for which the clean limit no longer applies. In this regime we find a cooling power which is accurately described qualitatively, but not quantitatively, by a model which predicts the emission of very high energy phonons through a disorder-assisted mechanism.
\end{abstract}
\maketitle
\section{\label{sec:Intro}Introduction}
The potential applications for graphene as a highly sensitive photon detector have driven substantial interest in determining the thermal conductance of graphene at low temperatures \cite{Fong2012a,Karasik2014b}. In addition to the scientific value of a greater understanding of electron-phonon coupling in graphene, knowledge of this important physical process is critical to determining the theoretical performance of highly sensitive graphene-based photon detectors. If the thermal conductance of an ultra-sensitive graphene-based photon detector is too large, the detector will cool off too quickly to allow for accurate photon detection \cite{McKitterick2013,McKitterick2015}. To date, thermal measurements of graphene have used Johnson noise thermometry \cite{Fong2012a,Fong2013,Betz2012a,Betz2012}, the temperature dependent resistance of a superconducting tunnel barrier \cite{Vora2014} or supercurrent hysteresis \cite{Borzenets} to measure the thermal conductance of graphene as a function of dissipated power, which determines the electron temperature. However, due to wide variability between samples, even within a single study, it has been difficult to form general conclusions about the phonon cooling pathway in graphene. In addition, the measurements are challenging. Typically the sample resistance is large, the contacts can add extra resistance and provide an additional cooling pathway, and the signals in these measurements are small. We discuss these issues below in detail. 

In metallic thin films (film thickness $\sim 10$~nm to 1~$\mu$m), the electron-phonon cooling power typically takes the form $P=V\Sigma\left(T^\delta - T_0^\delta \right)$, where $V$ is the device volume, $\Sigma$ is a coefficient that describes the electron-phonon coupling strength, and $\delta$ is a parameter that varies from $4-6$, dependent upon the amount of disorder in the system \cite{Karvonen2004}. Similarly, in monolayer graphene the electron-phonon cooling power and thermal conductance, $G_\mathrm{ep}$, depend on the level of disorder, screening, and the temperature of the electron system. We discuss the phonon cooling in the applicable theories, and then present our experiments on two monolayer graphene samples. 

By accounting for the microwave losses associated with substrate electrons, we are able to accurately measure the cooling of the samples. We find that, at low temperatures, the cooling in both samples can be described by a combination of hot electron out-diffusion and phonon emission. By tuning the carrier density of the samples and measuring devices with areas that differ by an order of magnitude, we are able to definitively identify the phonon cooling pathway. We find $G_\mathrm{ep}$ to be consistent with the behavior expected for a pure monolayer graphene sheet with little disorder.

\section{Theory of Phonon Cooling}
\subsection{Clean limit}
The cooling power in the limit of large electron-impurity mean free path $\ell_\mathrm{mfp}$ (taken to be infinite) follows a power law form
\begin{equation}
\label{eq:Plaw}
P=\Sigma(T^\delta-T_0^\delta)
\end{equation}
and is due to the emission of longitudinal acoustic (LA) phonons into the graphene lattice \cite{Viljas2010}. In Eq.~\ref{eq:Plaw}, $T$ and $T_0$ are the temperatures of the electron system and the lattice, respectively. The forms of the coefficient $\Sigma$ and exponent $\delta$ depend on the temperature of the electron system, with a crossover temperature of order the Bloch-Gruneisen temperature $T_\mathrm{BG}=2s\sqrt{\pi n}/k_\mathrm{B}$ \cite{Voutilainen2011},
where $s=2\times 10^4~$m/s is the speed of sound in graphene, $n$ is the carrier density in graphene, and $k_\mathrm{B}$ is Boltzmann's constant. For $n=10^{12}~\mathrm{cm^{-2}}$, as is typical for samples on SiO$_2$, $T_\mathrm{BG}=54~$K. At temperatures $T\ll T_\mathrm{BG}$,
\begin{equation}
\label{eq:elphCleanLim}
P=A\Sigma_1 (T^4-T_0^4),
\end{equation}
where
\begin{equation}
\Sigma_1=\frac{\pi^2 D^2|E_\mathrm{F}|k_\mathrm{B}^4}{15\rho_\mathrm{M}\hbar^5 v_\mathrm{F}^3s^3}.
\end{equation}
Here, $\rho_\mathrm{M}$ is the mass density of graphene, $v_\mathrm{F}=10^6$~m/s is the Fermi velocity in graphene, $E_\mathrm{F}=\hbar v_\mathrm{F}\sqrt{\pi n}$ is the Fermi energy, and $D$ is the deformation potential of graphene. The deformation potential is a measure of the strength of electron-phonon coupling and has been studied theoretically \cite{Kaasbjerg2012,Borysenko2010,Li2010} and experimentally \cite{Graham2012,Voutilainen2011,Fong2013,Betz2012a}. Calculations find $D$ ranges from 2 to 70 eV, with theoretical predictions ranging from approximately 5 to 13~eV. 

For $T\gg T_\mathrm{BG}$, the cooling power for typical devices, with $E_\mathrm{F}\gg k_\mathrm{B}T$, is given by
\begin{equation}
\label{eq:elphCleanLimhighT}
P=g_1A(T-T_0),
\end{equation}
where
\begin{equation}
g_1=\frac{D^2 E_\mathrm{F}^4 k_\mathrm{B}}{2\pi\rho_\mathrm{M}\hbar^5 v_\mathrm{F}^6}.
\end{equation}
For very low carrier densities or operation at high $T$, where $E_\mathrm{F}<k_\mathrm{B}T$,
\begin{equation}
g_1=T^4\frac{7\pi ^3 k_\mathrm{B}^5 D^2}{30\rho_\mathrm{M}\hbar^5 v_\mathrm{F}^6}.
\end{equation}
This is not a regime we access in our measurements, as $E_\mathrm{F}$ is approximately equal to $k_\mathrm{B}T$ at $T=1000$~K at a carrier density of $n=10^{12}~\mathrm{cm^{-2}}$. On the disordered SiO$_2$ substrates which are used here, much lower carrier densities are difficult to achieve \cite{Chen2008} and temperatures above 1000~K are inaccessible.

\subsection{Disorder-assisted scattering}
The introduction of disorder (by means of a finite mean free path) has ramifications at both low temperatures and high temperatures. Theoretical \\
 \cite{Chen2012} find a different form of electron-phonon cooling power than given by Eq.~\ref{eq:elphCleanLim} at low temperatures. It is predicted that at temperatures below a crossover temperature $T_\mathrm{x}$, the electron-phonon coupling is enhanced and is larger than the value given by Eq.~\ref{eq:elphCleanLim}:
\begin{equation}
\label{eq:difflimLowT}
P=A\Sigma_2 (T^3-T_0^3),
\end{equation}
where
\begin{equation}
\Sigma_2=\frac{2\zeta(3)D^2|E_\mathrm{F}|k_\mathrm{B}^3}{\pi^2 \rho_\mathrm{M}\hbar^4v_F^3s^2\ell_\mathrm{mfp}}
\end{equation}
and $\zeta(n)$ is the Reimann zeta function; $\zeta(3)\approx 1.2$. The crossover temperature $T_\mathrm{x}$ is the temperature for which the expressions in Eqs.~\ref{eq:elphCleanLim}~and~\ref{eq:difflimLowT} are equal. This temperature is given by
\begin{equation}
\label{eq:tdis}
T_\mathrm{x}=\frac{30 \hbar s\zeta(3)}{\pi^4 k_\mathrm{B}\ell_\mathrm{mfp}}.
\end{equation}
Using a Drude model, we find a lower bound of $\ell_\mathrm{mfp}=50$~nm for our samples, which sets an upper bound of $T_\mathrm{x}\approx 1$~K.

At higher temperatures, above $T_\mathrm{BG}$, a different form of electron-phonon scattering is predicted \cite{Song2011} in which disorder in the graphene allows large momentum phonons to be emitted from the electron system. These large momentum scattering events dissipate energy of order $k_\mathrm{B}T$ and are coined ``supercollisions." In typical momentum-conserving scattering, the momentum of scattered phonons is constrained by the Fermi surface of the graphene (a circle in $k$-space of radius $|E_\mathrm{F}|/\hbar v_\mathrm{F}$). However, supercollisions allow phonons with much larger momenta to be emitted, with an accompanying recoil phonon which allows the net energy exchange to be momentum-conserving. The result is that the cooling power due to supercollisions for $T>T_\mathrm{BG}$ is predicted to be exactly half the cooling power predicted for low temperature disorder-assisted scattering:
\begin{equation}
\label{eq:supercoll}
P=\frac{\Sigma_2}{2}A(T^3-T_0^3).
\end{equation}

In order to predict the electron-phonon thermal conductance in graphene, it is critical to clearly state the assumptions made about the presence of disorder and the temperature regime in which the device is operating. Moreover, measurement of $G_\mathrm{ep}$ is nontrivial. Though several groups have made measurements of graphene's thermal properties at cryogenic temperatures \cite{Fong2012a,Betz2012a,Borzenets}, there is considerable disagreement between experiments about the functional form of $G_\mathrm{ep}$ as well as the magnitude of $D$ \cite{Viljas2010}. In our measurements, we study very large area graphene samples to emphasize $G_\mathrm{ep}$, which is proportional to the graphene area. This enables us to clearly separate the electron-phonon cooling in our samples from other cooling pathways (notably, hot-electron out-diffusion). In addition, by looking at similar graphene samples with substantially different area, we are able to accurately probe the area dependence of the electron-phonon cooling channel.


\section{\label{sec:Fab}Device fabrication}
The samples used in our measurements are prepared from CVD-grown graphene purchased from ACS Material \cite{ACS2014}. The graphene is grown using copper foil as a catalyst. Prior to purchase, the graphene was transferred to an oxidized doped silicon substrate by coating the graphene sample with PMMA and etching away the copper foil. It has been shown that this process can leave PMMA residue on the surface of the graphene \cite{Her2013}. However, we found that the ability to lithograph large graphene areas to emphasize $G_\mathrm{ep}$ and reduce contact resistance outweighed the negative aspects of having possible contaminants. The doped Si substrate with a room temperature resistivity of $\rho=1~\Omega\cdot \mathrm{cm}$ allows for the carrier density of the graphene to be controlled in situ by using an electrostatic gate voltage to capacitively induce either electrons or holes. However, the doped substrate can complicate the measurement of device temperature (discussed in Section~\ref{sec:Mes}).

After purchase, the samples are fabricated in a multi-step process. Using electron-beam lithography, we define areas of the graphene sheet which we then remove with an oxygen etch. What remains are graphene sheets of width 100~$\mu$m and lengths ranging from 10 to 200~$\mu$m. Then, we again use the electron beam patterning, this time to define the contact and lead structure. We deposit a Pd/Al bilayer with thicknesses 5~nm/50~nm to contact the graphene. Palladium is used to form Ohmic contact to the graphene \cite{Xia2011,Zhong2014} and aluminum was used to realize superconducting contacts with a $T_\mathrm{c}\approx 1.2$~K. Superconducting contacts were desired to suppress the out-diffusion of hot electrons, another potential source of thermal conductance to the bath.

\section{\label{sec:Mes}Measurements}

The measurements for this study were performed in an Oxford Triton 200 cryogen-free dilution refrigerator. With no cabling, this refrigerator can achieve a base temperature of 8~mK.

\begin{figure}[htbp]
\centering
\includegraphics[width=.5\columnwidth]{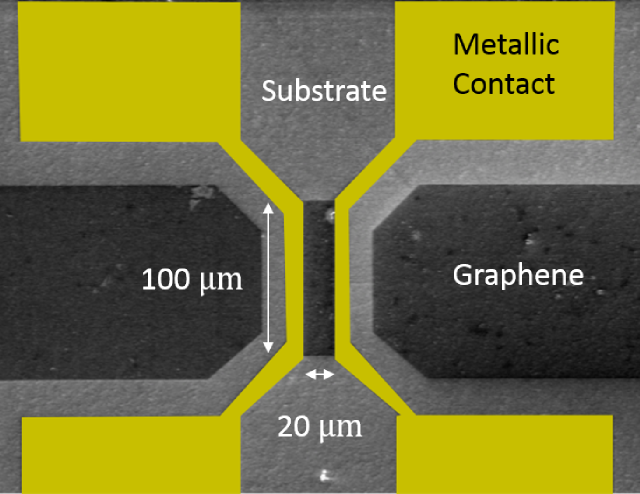}
\caption{Scanning electron microscope image of a typical graphene sample. The metallic leads are false colored in yellow. The dark regions are areas of graphene which were not etched away and the remaining light area is the SiO$_2$ substrate. Note that the channel length $L$ for the sample in the image is 20~$\mu$m. For the samples measured, $L$ is 10 or 100~$\mu$m. For both the sample shown in this figure and those measured, the channel width $W$ is $100~\mu$m.}
\label{fig:FabDevice}
\end{figure}

\begin{table}
\centering
\begin{tabular}{cccccc}
\toprule
Sample & $L$ ($\mathrm{\mu m}$)\hspace{5mm}& $W~(\mu\mathrm{m})$\hspace{3mm} & $R$~($\Omega$)\hspace{3mm} & $R_\square$~($\Omega$)\hspace{3mm} & $D$ (eV) \\\colrule
G1 & 10 & 100 & 90 & 900 & 12.0\\
G2 & 100 & 100 & 840 & 840 & 10.5\\\botrule
\end{tabular}
\caption{Comparison between the shorter and longer graphene samples (G1 and G2, respectively) at a gate voltage of $V_\mathrm{g}\approx-32$~V. The values of $D$ were obtained by individual fitting the thermal conductance as a function of $T$ using Eq.~\ref{eq:DiffEq} for $T<T_\mathrm{BG}/4$ at each gate voltage, Fig.~\ref{fig:Comparison} below.}
\label{table2}
\end{table}

\begin{figure}[htbp]
	\includegraphics[width=\columnwidth]{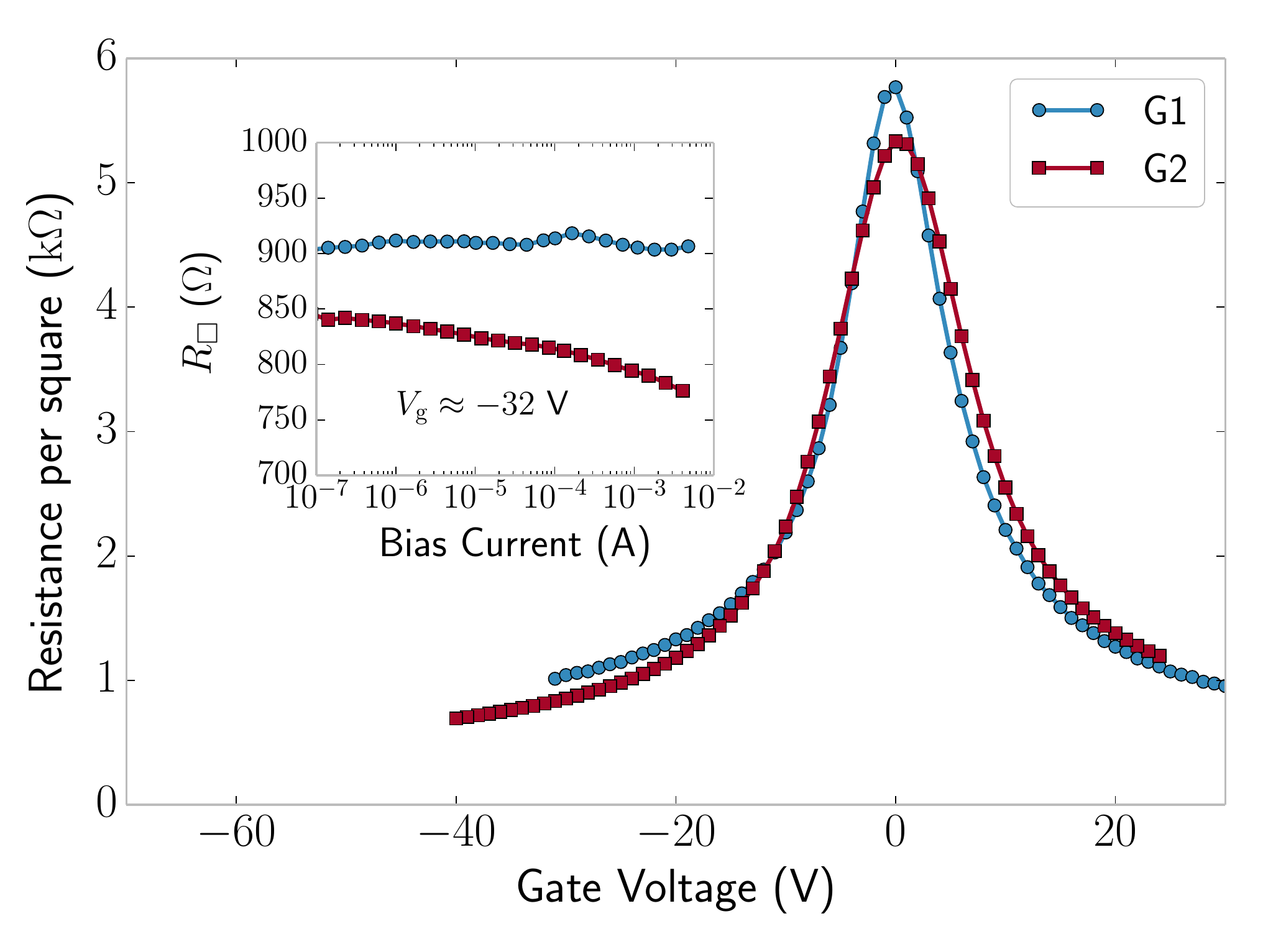}
	\caption{\label{fig:gates} Resistance per square, $R_\square$, as a function of gate voltage for sample G1 ($L=10~\mu$m) and sample G2 ($L=100~\mu$m) at a bias current of $I=1~\mu$A. The gate voltages of each curve are offset so that the resistance peak occurs at $V_\mathrm{g}=0$. The offsets are $-1.5~$V and $-8$~V for samples G1 and G2, respectively. Inset: Resistance per square as a function of bias current at $V_\mathrm{g}\approx-32$~V. Data were taken at $T_0<100$~mK.}
\end{figure}

We present the thermal measurements of the two samples(typical geometry shown in Fig.~\ref{fig:FabDevice}) with device properties given in Table~\ref{table2} from the same graphene growth, both with channel width $W=100~\mu$m. The first has length $L=10~\mu$m between contacts (sample G1) and the other has a channel length of 100~$\mu$m (sample G2). Measuring these two devices in a single cool down of the refrigerator allowed us to establish how much contact resistance is present and study the area-dependence of thermal conductance. In Fig.~\ref{fig:gates} we plot the resistance per square ($R_\square$) of both devices as a function of gate voltage. We define $R_\square=R(W/L)$ where $R$ is the measured resistance. The measured resistance can have a contribution due to the series contact resistance if present, but none due to the Al/Pd leads for $T<T_\mathrm{c}$. At negligible contact resistance, $R$ should scale with length for fixed $W$. Indeed, we find that $R_\square$ is approximately the same for the two samples. At gate voltages $|V_\mathrm{g}|>10$~V, far from the charge neutrality point (CNP), we calculate a mobility of approximately 3500~$\mathrm{cm/s^2}$ for both electrons and holes, which is consistent with high quality CVD-grown graphene on SiO$_2$ \cite{Gannett2011}. We find that the resistance is relatively insensitive to bias current over several orders of magnitude of current. The comparison of $R_\square$ at $V_\mathrm{g}=-32$~V for samples G1 and sample G2 indicates that the contact resistance is small compared to the total device resistance. In addition, analysis of the conductivities as a function of gate voltage \cite{Du2008} yields residual root-mean-square carrier densities of $n\approx 2.5\times 10^{11}~\mathrm{cm^{-2}}$ and $3\times 10^{11}~\mathrm{cm^{-2}}$ at the resistance peak for samples G1 and G2, respectively. This is due to carrier puddling arising from external electric fields \cite{Martin2007}.

\begin{figure}[htbp]
\includegraphics[width=\columnwidth]{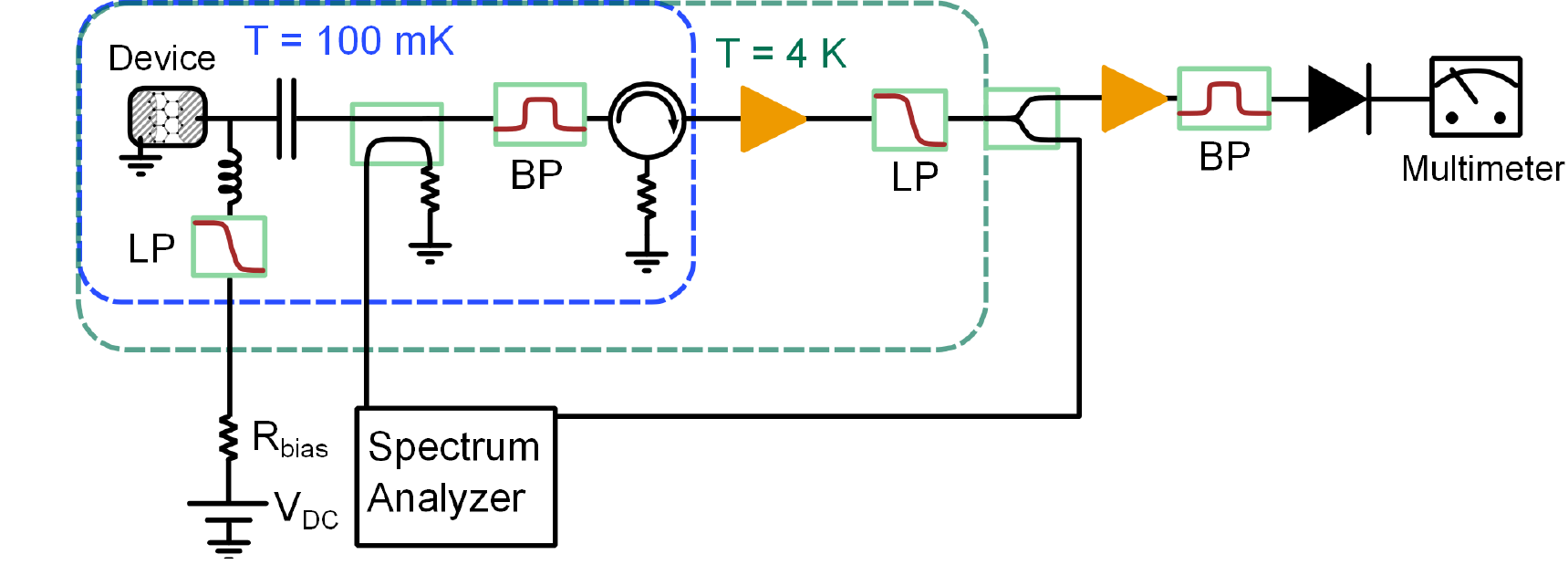}
\caption{\label{fig:apparatus} Apparatus for Johnson noise measurements. The metallic leads contacting the device are represented by the hatched regions. The low-pass and band-pass filters are indicated by LP and BP (with center frequency $f=1.3$~GHz), respectively. The bias-T at base is represented by a capacitor and inductor. The rf signal from the spectrum analyzer is attenuated by 40~dB before coupling to the device and is necessary for the reflectometry measurement described in the text.}
\end{figure}

To study the thermal properties of the graphene, a constant power is applied to the electron system and the emitted Johnson noise of the system is measured to determine the resulting change in electron temperature. These steady state measurements allow us to probe the cooling pathways of the device. We use a Yokagawa 7651 voltage source and a large bias resistor to apply a dc or low frequency on/off current of amplitude $I$ to the graphene (see Fig.~\ref{fig:apparatus}). The current heats the electrons in the graphene with power $P=I^2R$. This heating power raises the steady state electron temperature of the graphene above the stage temperature $T_0$. This change in electron temperature results in a change in emitted Johnson noise power into a matched load equal to $k_\mathrm{B}B\Delta T$, where $\Delta T=T-T_0$. This result assumes that the emitted noise is measured at a sufficiently low frequency $f$ so that $hf\ll k_\mathrm{B}T$, as is the case in our study with $f=1.3$~GHz, and that the electron temperature is constant across the device. In the case of finite thermal conductance from charges diffusing out the leads, the electron temperature as a function of position $T(x)$ is not constant, so the measurement of Johnson noise probes the average electron temperature of the graphene $\bar{T}$:
\begin{equation}
\label{eq:avtemp}
\bar{T}=\frac{1}{L}\int\limits_{0}^{L}T(x)dx
\end{equation}
and $\Delta T=\bar{T}-T_0$.

The Johnson noise signal is rectified to produce a dc voltage using a zero bias Schottky diode. The change in diode voltage is given by $\Delta V_\mathrm{diode}=\kappa \Delta T$, where $\kappa$ is a coupling constant representing the amplification of the 50~$\Omega$ microwave output system. In order to accurately measure the average temperature of the electron system, it is thus critical to determine the value of $\kappa$. This is one of the important calibration procedures necessitated by the use of a commercially-supplied doped Si substrate, which is weakly electrically conducting. 

\begin{figure}[htbp]
	\includegraphics[width=\columnwidth]{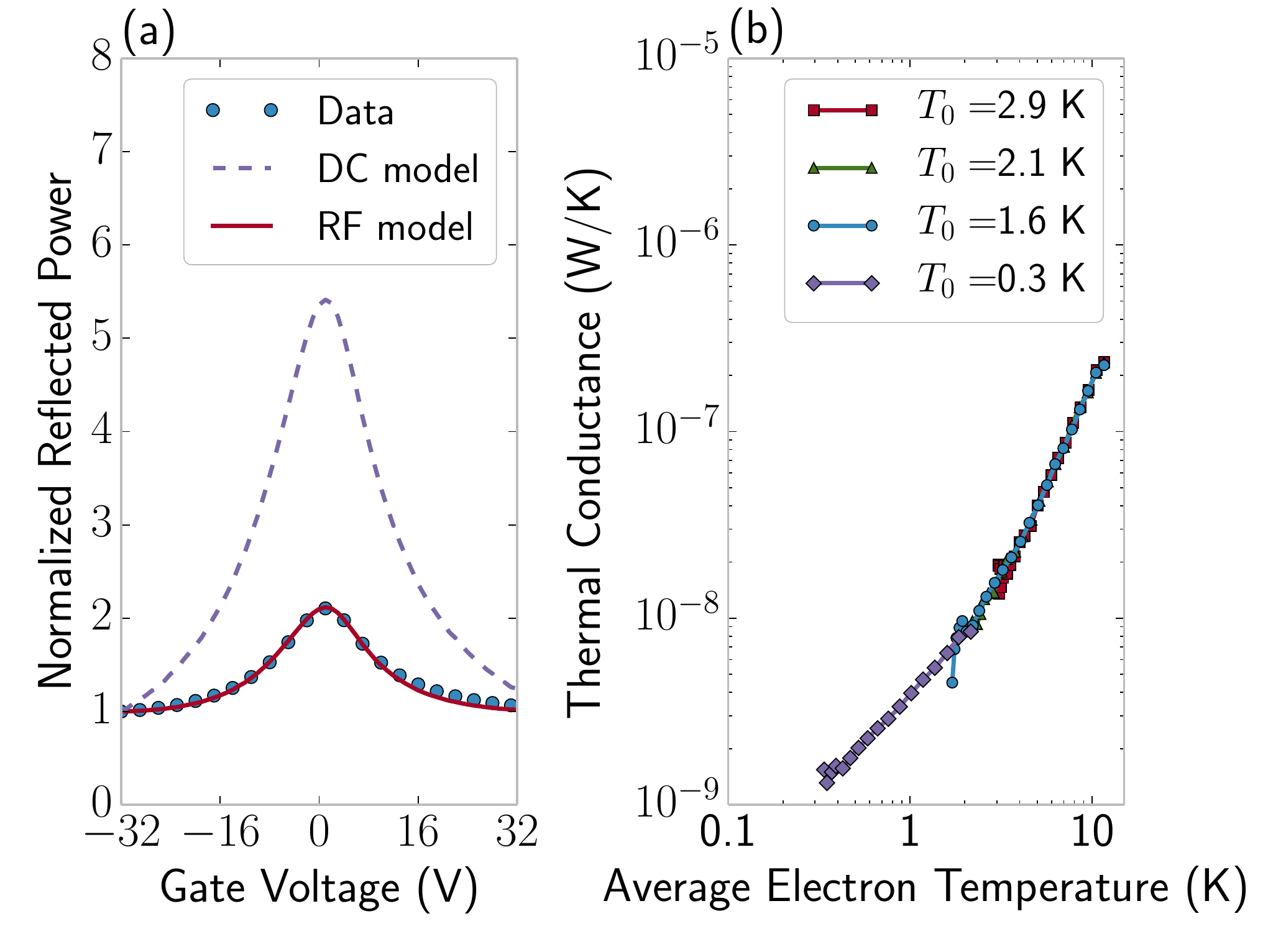}
	\includegraphics[width=.6\columnwidth]{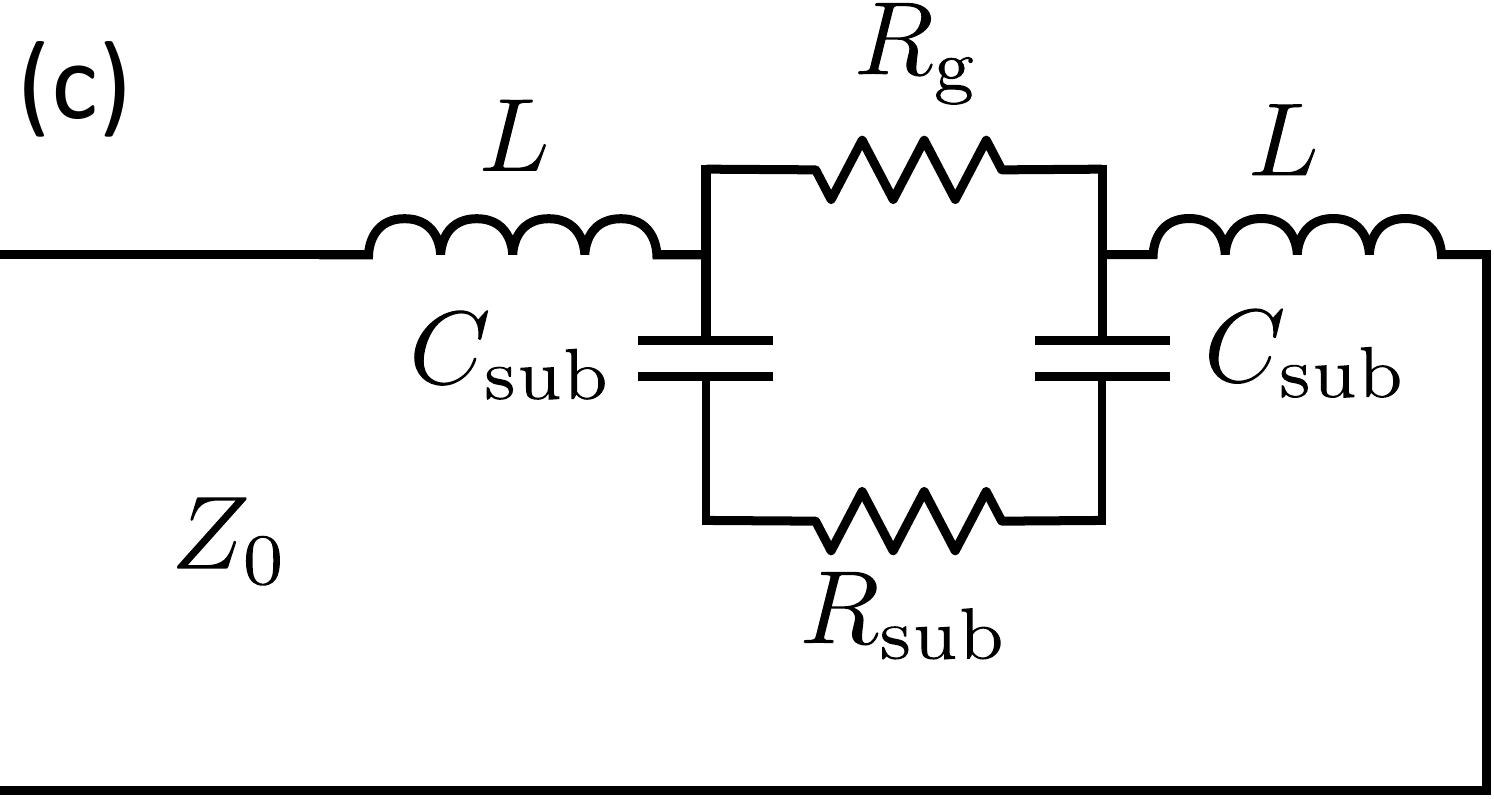}
	\caption{\label{fig:reflpow} Characterization and calibration of Sample G1. (a) Reflected power from Sample G1 as a function of gate voltage. The DC model gives the reflected power that would be expected from calculating the reflection coefficient using the device resistance. The LCR model instead calculates reflection using the circuit model of (c), with $L=2.5~$nH, $C_\mathrm{sub}=10$~pF, $R_\mathrm{sub}=290~\Omega$, and $R_\mathrm{g}$ equal to the internal graphene resistance. The capacitance was calculated from the area of the lead structure, the inductance was estimated from wirebond length, and the substrate resistance was adjusted to provide a good fit to the data. At $V_\mathrm{g}=-32~$V, the circuit model yield a power reflection of $\Gamma^2=0.15$, while the DC model predicts $\Gamma^2=0.09$. Thus, the device is well matched when gated far from the CNP. (b) Calculated thermal conductance as a function of average electron temperature for several base temperatures. To determine the coupling, $\kappa$ is chosen (explained in the text) to align the thermal conductance curves (so that $G(\bar{T})$ is independent of $T_0$).}
\end{figure}

To better understand the microwave coupling to the device, we performed reflectometry measurements using a spectrum analyzer. In Fig.~\ref{fig:reflpow}a the normalized reflected power as a function of gate voltage is plotted for sample G1. For comparison, the expected normalized reflected power is also plotted. Here we consider the impedance mismatch between the 50~$\Omega$ microwave line and an equivalent resistor with the dc resistance of the graphene sample. Although the rf impedance at 1~GHz of the graphene itself is approximately equal to its dc resistance \cite{Horng2011}, the data deviates substantially from the calculation seen in Fig.~\ref{fig:reflpow}a, where the dc resistance is used to calculate the circuit rf impedance. We believe the discrepancy is not due to the graphene, but instead arises from charge carriers in the doped Si substrate capacitively coupling through the large contact pads. Using the very simple circuit model shown in Fig.~\ref{fig:reflpow}c where the carriers in the substrate provide a parallel resistance, we can approximately replicate the observed microwave behavior, with the values of the lumped circuit elements given in the figure caption.

The coupling of substrate charges to the Al/Pd leads means that the carriers in the substrate also emit Johnson noise, which is measured at the output. As a result, a calibration based on the diode output voltage as a function of stage temperature would yield a slope that is larger than $\kappa$. Instead, we determine the coupling by measuring the diode voltage as a function of DC input power at different stage temperatures and requiring that the $G(T)=dP/d\bar{T}$ be independent of the stage temperature at which the measurement was performed (see Fig.~\ref{fig:reflpow}b). It is important to note that this method is only exactly correct in the case of uniform electron temperature across the graphene device. In our system, where we observe a non-uniform temperature distribution due to carrier out-diffusion, we find from simulations that this calibration technique can understate the value of $\kappa$ by 10 to 15\%. We account for this effect in the data we present in Figs.~\ref{fig:GLowT},\ref{fig:Comparison}, and \ref{fig:HighT}.


\subsection{\label{sec:CModel}Cooling model}
In analyzing the thermal properties of the samples, the heat diffusion equation is used to model to cooling power of the system:
\begin{equation}
\label{eq:DiffEq}
I^2r=p_\mathrm{ep}(x)-\frac{\partial}{\partial x}\left(g(x)\frac{\partial T(x)}{\partial x}\right),
\end{equation}
where $r$ is the resistance per unit length, $p_\mathrm{ep}(x)$ is the electron-phonon cooling power per unit length, and $g(x)=\mathcal{L}T/R_\square$ is the thermal conductance from carrier diffusion. The appropriate form of $p_\mathrm{ep}(x)$ depends both on the electron temperature and on the level of disorder in the system (i.e., $\ell_\mathrm{mfp}$). For example, in the low temperature, clean limit, $p_\mathrm{ep}(x) = W\Sigma_1\left(T(x)^4-T_0^4\right)$. After solving for $T(x)$ at a given current, the average electron temperature of the system $\bar{T}$, can be found by integrating over the entire length, as per Eq.~\ref{eq:avtemp}.

This average temperature is what is measured using the Johnson noise method. Note that this calculation is correct only if the resistance is approximately temperature independent, which is very close to correct for our samples (Fig.~\ref{fig:gates}).

The boundary condition used to solve Eq.~\ref{eq:DiffEq} is that $T(x)=T_0$ at both ends of the device, as there was strong evidence of diffusion cooling of hot electrons in sample G1. Because sample G2 has a resistance and area that is approximately 10 times larger than that of sample G1, diffusion cooling is not evident in our experiments. The measurements of $\bar{T}$ were performed over a large range of excitation powers ($I=0.1-1000~\mu$A), so we access electron temperatures above and below $T_\mathrm{BG}$ in our measurement. We model these two regimes separately, first focusing on low temperatures, below $T_\mathrm{BG}/4$.
\subsection{\label{sec:LowT}Low temperature cooling}
\begin{figure}[htbp]
\includegraphics[width=\columnwidth]{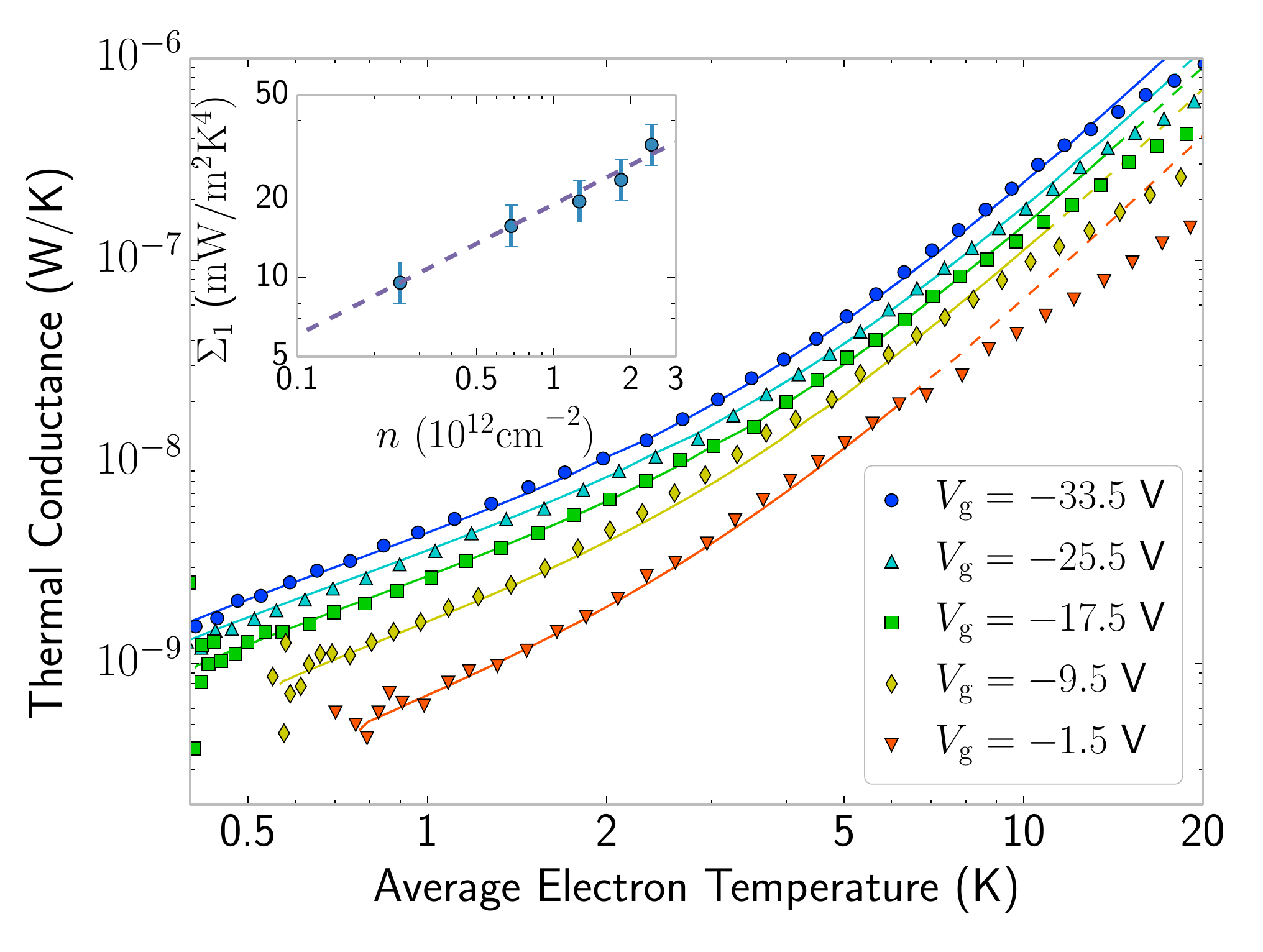}
\caption{\label{fig:GLowT} Thermal conductance, $G=dP/d\bar{T}$ of sample G1. The solid line is a calculation of thermal conductance for $T<T_\mathrm{BG}/4$  using the device parameters given in Table~\ref{table} and Eqs.~\ref{eq:elphCleanLim}~and~\ref{eq:DiffEq}. The same calculation is shown by a dashed line for $T>T_\mathrm{BG}/4$. Inset: calculation of $\Sigma_1$ using $n$ calculated from $V_\mathrm{g}$ and measured values of $D$ (given in Table~\ref{table}). The dashed line shows the value $\Sigma_1$ as a function of $n$ for $D=11.5$~eV, demonstrating the expected $\Sigma_1\propto\sqrt{n}$ dependence. The uncertainty at each of the data points for $\Sigma_1$ arises from the calibration procedure at each gate voltage. There is an additional overall uncertainty of scale for $\Sigma_1$ (of $\pm 20\%$) due to uncertainty in the stage thermometer calibration.}
\end{figure}


In Fig.~\ref{fig:GLowT}, we plot the thermal conductance of sample G1 for $T<20$~K as a function of the average electron temperature $\bar{T}$ for multiple gate voltages (in the hole doped region). The solid lines are plotted for $T<T_\mathrm{BG}/4$ for each density. These plots are a calculation of the effective thermal conductance, $G=dP/d\bar{T}$, using the temperature distribution determined by Eq.~\ref{eq:DiffEq}. At low temperatures, for each gate voltage, the thermal conductance of this sample is approximately equal to the value predicted for the out-diffusion of hot electrons to the contacts a one-dimensional wire: $G_\mathrm{diff}\approx13\mathcal{L}\bar{T}/R$, where $\mathcal{L}$ is the Lorentz number \cite{Kittel2004,Prober1993}. In our measurements, we find $\mathcal{L}=1.18\times \mathcal{L_\mathrm{theory}}$, where $\mathcal{L_\mathrm{theory}}=2.45\times 10^{-8}~\mathrm{W\Omega K^{-2}}$.

\begin{table}
	\centering
	\begin{tabular}{ccccc}
		\toprule
		$V_\mathrm{g}$ (V)\hspace{5mm} & $n$ ($10^{11}~\mathrm{cm}^{-2}$)\hspace{5mm}& $T_\mathrm{BG}$~(K)\hspace{5mm} & $R$ ($\Omega$)\hspace{5mm} & $D$ (eV) \\\colrule
		-33.5 & 24.1 & 84 & 90 & 12.0\\
		-25.5 & 18.4 & 73 & 110 & 11.0\\
		-17.5 & 12.6 & 60 & 150 & 11.0\\
		-9.5 & 6.8 & 45 & 250 & 11.5\\
		-3.5 & 2.5 & 27 & 580 & 11.5\\\botrule
	\end{tabular}
	\caption{Device properties of sample G1 ($L=10~\mu$m). The values of $D$ were obtained by individually fitting the thermal conductance as a function of $T$ using Eq.~\ref{eq:DiffEq} for $T<T_\mathrm{BG}/4$ at each gate voltage, Fig.~\ref{fig:GLowT}.}
	\label{table}
\end{table}

At temperatures above a few Kelvin, the cooling is increasingly dominated by the emission of phonons, and we use the clean limit form of $p_\mathrm{ep}$ to calculate the electron temperature. This determines $\Sigma_1$. The only free parameter in determining the thermal conductance is the deformation potential $D$, which appears in Eq.~\ref{eq:elphCleanLim}. For each gate voltage, the value of $D$ was independently determined from $\Sigma_1$ by fitting to the relevant data points, and the plotted lines represent the best fit. The values of $D$ found range from $11$ to $12\pm2.0$~eV and are presented along with other relevant physical parameters in Table~\ref{table}. With these values of $D$, we have calculated $\Sigma$ as a function of $n$ and have plotted these results in the inset of Fig.~\ref{fig:GLowT}. The dashed line in the inset represents the anticipated $n$ dependence of $\Sigma_1\propto \sqrt{n}$ for a deformation potential of $D=11.5$~eV. 

\begin{table*}
	\centering
	\begin{tabular}{cccccc}
		\toprule
		Study & Regime\hspace{5mm}& Substrate\hspace{5mm}& $A~(\mu\mathrm{m}^2)$\hspace{5mm} & $R$ (k$\Omega$)\hspace{5mm} & $D$ (eV) \\\colrule
		Betz \emph{et al.} \cite{Betz2012a} & Supercollisions & BN& 6.2 & 1.5 & 70\\
		Betz \emph{et al.} \cite{Betz2012} & Clean limit & BN& 6,13 & 1,3 & 4,2\\
		Fong \emph{et al.} \cite{Fong2012a} & Clean limit & SiO$_2$/Si& 102 & 30 & 33\\
		Fong \emph{et al.} \cite{Fong2013} & Low-$T$ disorder & SiO$_2$/Si& 25,55 & 1.5,5 & 51,23\\
		Theory \cite{Kaasbjerg2012,Borysenko2010,Li2010}& -- & -- & -- & -- & 5-13\\
		This study & Clean limit & SiO$_2$/Si & $10^4,10^5$ & 0.09,0.84 & 12,11\\\botrule
	\end{tabular}
	\caption{Summary of experimental results which studied the low-temperature electron-phonon coupling in graphene using Johnson noise thermometry. Individual samples within a study are separated by commas.}
	\label{otherstud}
\end{table*}
We also tested the disorder-assisted scattering model, with $p_\mathrm{ep}$ given by Eq.~\ref{eq:difflimLowT}, in solving Eq.~\ref{eq:DiffEq}.  We found that this model did not agree with the data. We conclude that the data do not exhibit an electron-phonon cooling power with a $T^3$ dependence at temperatures below 20~K. This conclusion is consistent with measurements of device resistance, which indicate a disorder-limited mean free path of over 50~nm. For this value of $\ell_\mathrm{mfp}$, the crossover temperature $T_\mathrm{x}$, below which disorder-assisted scattering would be expected to play a role, is $T_\mathrm{x}\approx1$~K. For all gate voltages measured, the thermal conductance at these lowest temperatures ($T\lesssim 1$) is dominated by electron out-diffusion. Thus, we could not test the disordered limit in the present experiments

The clean limit calculation ceases to agree with the measured data at large temperatures, and the temperature at which this occurs is dependent upon the gate voltage. This is to be expected, as $T_\mathrm{BG}\propto\sqrt{n}$ so that $T_\mathrm{BG}/4\approx 11~$K at $V_\mathrm{g}=-9.5$, for example, whereas $T_\mathrm{BG}/4\approx 21~$K for $V_\mathrm{g}=-33.5$. Thus, the temperature range in which the low-temperature clean limit accurately describes $p_\mathrm{ep}$ is smaller for lower carrier densities. 

\begin{figure}[htbp]
\includegraphics[width=\columnwidth]{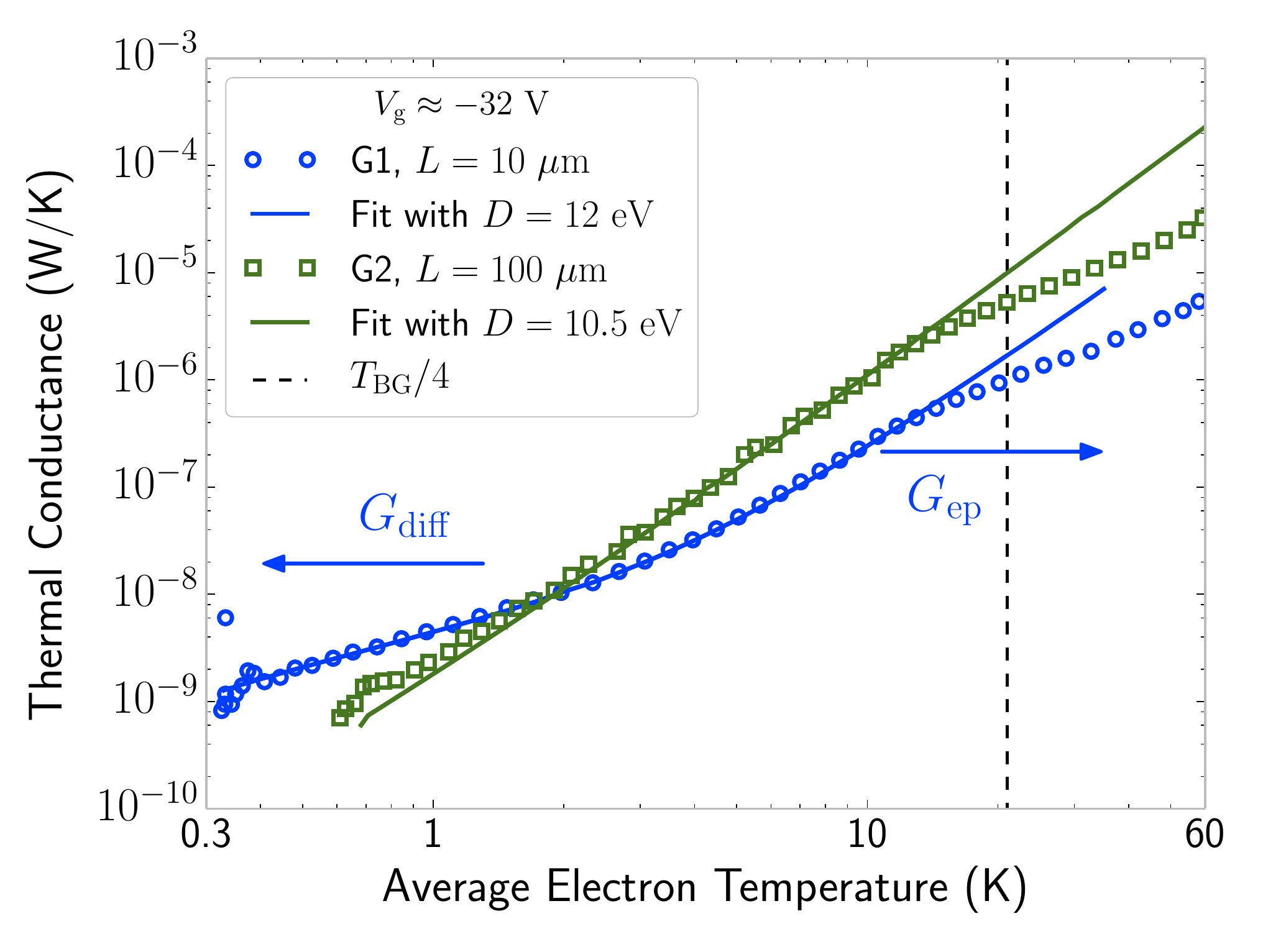}
\caption{\label{fig:Comparison} Plot of thermal conductance for samples G1 and G2. The solid lines are calculations of thermal conductance using Eq.~\ref{eq:DiffEq} in the low temperature clean limit, with $D=12$~eV and $D=10.5$~eV for samples G1 and G2, respectively. The dashed vertical line indicates $T_\mathrm{BG}/4\approx 21$~K. The temperature ranges over which either $G_\mathrm{diff}$ or $G_\mathrm{ep}$ is the dominant source of thermal conductance in sample G1 are indicated by the arrows in the figure. For sample G2, $G_\mathrm{ep}\gg G_\mathrm{diff}$ for all temperatures measured.}
\end{figure}

We tested the predicted area dependence of Eq.~\ref{eq:elphCleanLim} by comparing the thermal conductance of the shorter and longer samples (G1 and G2, respectively). In Fig.~\ref{fig:Comparison}, we plot the thermal conductance at high density, $V_\mathrm{g}\approx-32$~V, for both sample G1 and sample G2. At low temperatures, the total thermal conductance of sample G2 is less than that of sample G1 due to the approximately 10 times larger internal resistance of sample G2 (see Table~\ref{table2}). The increased length of G2 suppresses $G_\mathrm{diff}$ by a factor of 10, resulting in a much lower total thermal conductance below 1~K. However, at high temperatures, the thermal conductance of sample G2 is approximately 10 times larger than that of sample G1 as expected. The fits for samples G1 and G2 yield $D=12$~eV and $D=10.5$~eV, respectively, in good agreement and consistent within the measurement uncertainty.

\subsection{\label{sec:HighT}High temperature cooling}
\begin{figure}[htbp]
\includegraphics[width=\columnwidth]{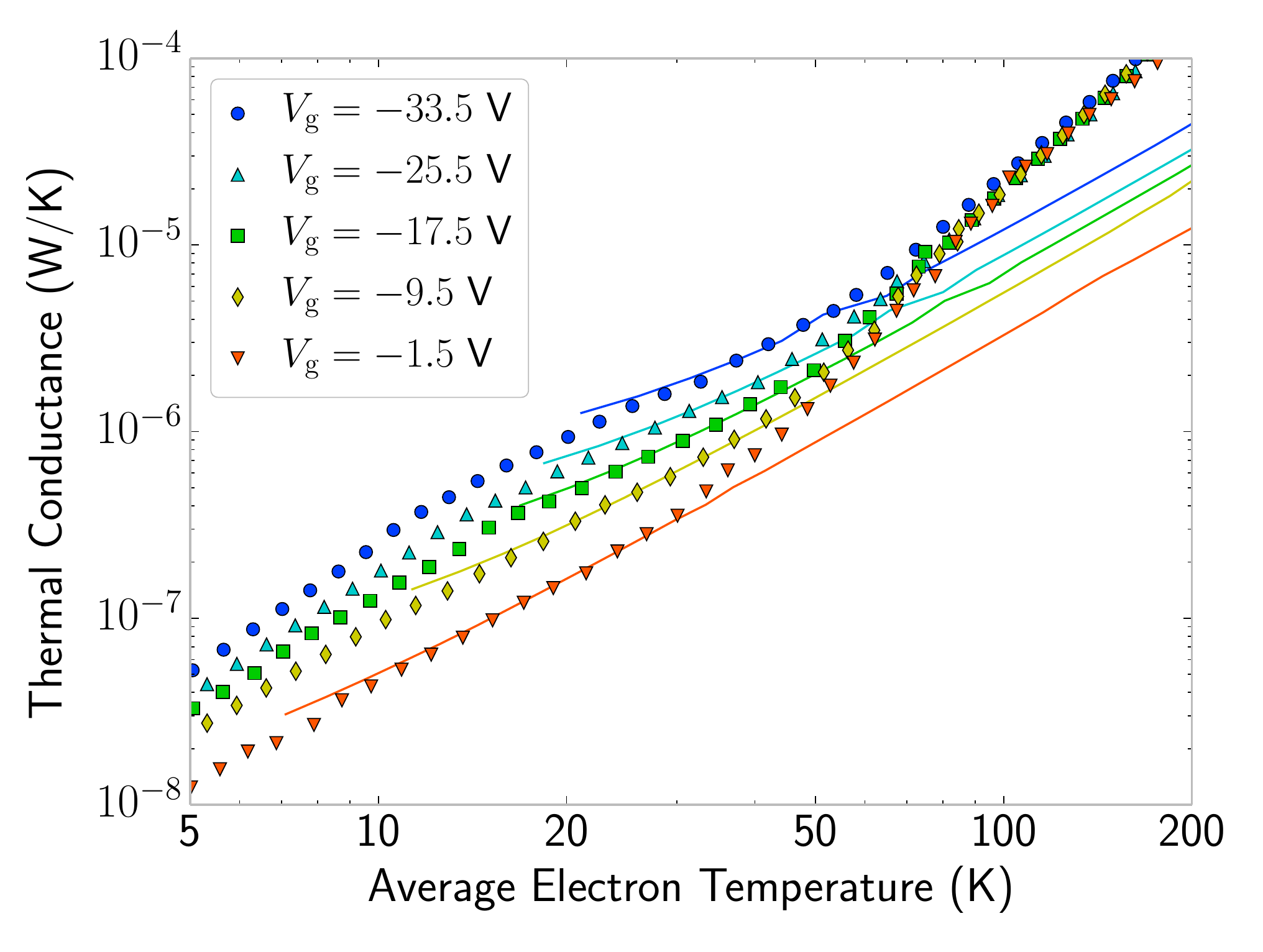}
\caption{\label{fig:HighT} Measured thermal conductance, $G=dP/d\bar{T}$, of sample G1. The solid line is a calculation of thermal conductance for $T>T_\mathrm{BG}/4$ using the device parameters given in Table~\ref{table}, $\ell_\mathrm{mfp}=3~$nm, and Eqs.~\ref{eq:elphCleanLimhighT},~\ref{eq:supercoll},~and~\ref{eq:DiffEq} This value of $\ell_\mathrm{mfp}$ is much smaller than expected, as discussed in the text.}
\end{figure}

At high bias currents, the electron temperature can exceed $T_\mathrm{BG}/4$ so that the electron-phonon cooling of the graphene can no longer be described by its low temperature limit. In Fig.~\ref{fig:HighT} we plot the thermal conductance of sample G1 at many gate voltages for $T>10~$K. Over a temperature range of approximately 15~K to 50~K, the thermal conductance follows an approximately $T^2$ power law, which suggests that supercollisions may contribute to $G_\mathrm{ep}$. As a result, we used a linear sum of Eqs.~\ref{eq:elphCleanLimhighT}~and~\ref{eq:supercoll} to model the electron phonon cooling power, using the values for $D$ that were extracted at lower temperatures. In these calculations, the only free parameter is the mean free path due to short range disorder potentials. 

We find that the best fit to the supercollision model is obtained with $\ell_\mathrm{mfp}=3\pm 1$~nm. 
This value for electron-impurity mean free path is quite low and is much lower than would be expected from the lower bound obtained by using the Drude model with the device resistance. Moreover, such a small value for $\ell_\mathrm{mfp}$ would suggest that at lower temperatures, disorder-assisted scattering would have a substantial contribution to the thermal conductance. Such behavior is not observed. Other groups using supercollisions to model their data have found very large values for the deformation potential ($D>50~$eV) by fixing $\ell_\mathrm{mfp}$ to an estimated value \cite{Betz2012a,Graham2012}. Because the coefficient $\Sigma_2$ in Eq.~\ref{eq:supercoll} is proportional to $D^2/\ell_\mathrm{mfp}$, both our results and those of previous studies find that attempting to explain the thermal conductance data with the supercollision model results in extracted physical parameters $D$ and/or $\ell_\mathrm{mfp}$ that are not within the expected range for those devices. 

At $T>50~$K, the thermal conductance appears to increase faster as a function of $T$ than would be expected for a $T^2$ dependence. Also, the curves asymptote to a value which is independent of gate voltage. This behavior is not anticipated from models of LA phonon scattering within the graphene, so we believe that this is evidence of the emission of optical phonons directly into the SiO$_2$ substrate \cite{Santavicca2010,Ong2012,Li2010}. At these high energies, the large potential energy drop over a typical electron mean free path might exceed the energy needed to emit a surface polar photon. 

\section{\label{sec:Disc}Discussion}
The Johnson noise emission measurements reported in the previous section probed the cooling processes of two graphene samples with areas which differed by a factor of 10. This made it possible to observe the dependence of cooling power on the device area which is consistent with theory and, in the case of the longer device, remove the effects of electron out-diffusion. The measurements of both the 10~$\mu$m long and of the 100~$\mu$m long sample found similar results for the deformation potential $D$, with measured values ranging from $10.5-12$~eV. Due to systematic uncertainties (primarily arising from the calibration and thermometry), there is an overall uncertainty in $D$ of approximately $\pm 20\%$. As a result, the bound that can reasonably be placed on the deformation potential from these measurements is $8.5-13.5$~eV (see~Table~\ref{table}). In Table~\ref{otherstud}, we compare this result to the extracted values of $D$ obtained from other studies, which show wide variation from sample to sample and little agreement with the theoretical predictions.

At low temperatures, the electron-phonon cooling is consistent with the clean limit for low electron temperatures ($T<20~$K for $n=2.4\times 10^{12}/\mathrm{cm^2}$), for which $G\propto T^3$ is predicted. We did not observe any behavior consistent with low temperature disorder-assisted scattering \cite{Chen2012} in the shorter samples, but the cooling pathway of hot electrons diffusing out the leads might obscure a deviation from the clean limit in the shorter sample, G1, at low temperatures if one were present. For the much longer sample, G2, a small deviation from $G\propto T^3$ was observed at very low temperatures for all gate voltages (see Fig.~\ref{fig:Comparison}). However, there is greater relative uncertainty in the graphene electron temperature below 1~K, which precludes drawing any quantitative conclusions for the electron-phonon cooling processes at these very low temperatures.

A interesting future study would be thermal measurements of a graphene sample with low electrical resistance, which was contacted with high $T_\mathrm{c}$ metallic contacts ($T_\mathrm{c}>9~$K) that were able to confine hot electrons and suppress $G_\mathrm{diff}$. A low sample resistance would be desirable to achieve good microwave coupling to the output lines and to significantly mitigate the effects of free carriers in the substrate. Clearly, an insulating substrate would be desirable for these measurements, with a separate gate electrode which couples to the sample at dc, but not at microwave frequencies. In addition, if the sample were suspended \cite{Du2008,Bolotin2008}, thermal conductance measurements at high dc bias would provide a useful test of whether optical surface phonons can explain the sharp upturn in thermal conductance observed for $T>50$~K. However, care must be taken to ensure that suspending the device does not generate a bottleneck for removal of excess phonons from the graphene lattice as this could complicate the interpretation of the data.

\section*{Acknowledgements}
This work was made possible by NSF Grant DMR-0907082, an IBM Faculty Grant, and by Yale University. Facilities use at Yale was supported by YINQE and NSF MRSEC DMR 1119826. In addition we want to acknowledge Prof. Michel Devoret, Prof. Michael Hatridge, and Faustin Carter for fruitful discussions on experimental design. Finally, Prof. Xu Du and Dr. Heli Vora provided invaluable guidance throughout this project.

{\footnotesize
\bibliography{Article}}

\end{document}